\documentclass[english,aps,preprint,notitlepage]{revtex4-1}
\usepackage[T1]{fontenc}
\usepackage[latin9]{inputenc}
\usepackage{babel}
\usepackage{textcomp}
\usepackage{amsmath}
\usepackage{amssymb}
\usepackage{graphicx}
\usepackage[unicode=true]
 {hyperref}

\makeatletter
\usepackage{caption,setspace}
\makeatother

\begin{document}

\preprint{Manuscript submitted to Nature Communications}

\title{Direct observations of a surface eigenmode of the dayside magnetopause}

\author{M.O.~Archer}
\email{m.archer@qmul.ac.uk}
\address{ School of Physics and Astronomy, Queen Mary University of London,
Mile End Road, London, E1 4NS, UK.}
\address{ Space and Atmospheric Physics Group, Department of Physics, Imperial
College London, South Kensington Campus, London, SW7 2AZ, UK.}

\author{H.~Hietala}
\address{ Department of Earth, Planetary and Space Sciences, University of
California, Los Angeles, 595 Charles Young Drive East, CA 90095-1567,
USA.}
\address{ Space Research Laboratory, Department of Physics and Astronomy, University
of Turku, 20500 Turku, Finland.}

\author{M.D.~Hartinger}
\address{ Space Science Institute, 4750 Walnut St Suite 205, Boulder, CO 80301,
USA.}
\address{ Electrical and Computer Engineering Department, Virginia Tech, Perry
St, Blacksburg, VA 24060, USA.}

\author{F.~Plaschke}
\address{ Space Research Institute, Austrian Academy of Sciences, Schmiedlstra{\ss}e
6, 8042 Graz, Austria.}

\author{V. Angelopoulos}
\affiliation{ Department of Earth, Planetary and Space Sciences, University of
California, Los Angeles, 595 Charles Young Drive East, CA 90095-1567,
USA.}

\maketitle

\section*{Abstract}

The abrupt boundary between a magnetosphere and the surrounding plasma,
the magnetopause, has long been known to support surface waves. It
was proposed that impulses acting on the boundary might lead to a
trapping of these waves on the dayside by the ionosphere, resulting
in a standing wave or eigenmode of the magnetopause surface. No direct
observational evidence of this has been found to date and searches
for indirect evidence have proved inconclusive, leading to speculation
that this mechanism might not occur. By using fortuitous multipoint
spacecraft observations during a rare isolated fast plasma jet impinging
on the boundary, here we show that the resulting magnetopause motion and
magnetospheric ultra-low frequency waves at well-defined frequencies
are in agreement with and can only be explained by the magnetopause
surface eigenmode. We therefore show through direct observations that
 this mechanism, which should impact upon the magnetospheric system
globally, does in fact occur.

\section*{Introduction}

Planetary magnetic fields act as obstacles to solar/stellar winds
with their interaction forming a well-defined region of space known
as a magnetosphere. The outer boundary of a magnetosphere, the magnetopause,
is arguably the most significant since it controls the flux of mass,
energy, and momentum both into and out of the system, with the boundary's
motion thus having wide ranging consequences. Magnetopause dynamics,
for example, can cause loss of relativistic radiation belt electrons
\cite{li97}; result in field-aligned currents directing energy to
the ionosphere \cite{haerendel13}; and launch numerous modes of magnetospheric
ultra-low frequency (ULF) waves \cite{plaschke16,hwang16} that themselves
transfer solar wind energy to radiation belt \cite{mann13}, auroral
\cite{mottez16}, and ionospheric regions \cite{rae07}. On timescales
greater than $\sim6\,\mathrm{min}$ Earth's magnetopause responds
quasistatically to upstream changes to maintain pressure balance \cite{glassmeier08}.
Simple models treating the dayside magnetopause as a driven damped
harmonic oscillator arrive at similar timescales \cite{smit68,freeman95,borve11}.
How the boundary reacts to changes over shorter timescales is not
fully understood.

It was proposed that plasma boundaries, including the dayside magnetopause,
may be able to trap impulsively excited surface wave energy forming
an eigenmode of the surface itself \cite{chen74}. The magnetopause
surface eigenmode (MSE) therefore constitutes a standing wave pattern
of the dayside magnetopause formed by the interference of surface
waves propagating both parallel and anti-parallel to the magnetospheric
magnetic field which reflect at the northern and southern ionospheres.
Its theory has been developed using ideal incompressible magnetohydrodynamics
(MHD) in a simplified box model, as depicted in Figure~\ref{fig:cartoon}a-c
along with expected polarisations (panels d-e) \cite{plaschke11}. The signature
of MSE within the magnetosphere should be a damped evanescent fast-mode
magnetosonic wave whose perturbations could significantly penetrate
the dayside magnetosphere \cite{archer15}. While this simple model
neglects many factors which might preclude the possibility of MSE,
global MHD simulations and applications of the theory to more representative
models suggest MSE should be possible at Earth with a fundamental
frequency typically less than $2\,\mathrm{mHz}$ \cite{hartinger15,archer15}.
The considerable variability of Earth's outer magnetosphere, however,
might suppress MSE's excitation efficiency \cite{pilipenko17}. The
simulations have largely confirmed the theorised structure and polarisations
of MSE but revealed that the relative phase of the field-aligned magnetic
field perturbations differed from the box model prediction by $50^{\circ}$
\cite{hartinger15}.

There exist numerous possible impulsive drivers of MSE including interplanetary
shocks \cite{villante16}, solar wind pressure pulses \cite{zuo15},
and antisunward plasma jets \cite{plaschke18}, all of which are known
to result in magnetopause dynamics and magnetospheric ULF waves in
general. However, no direct evidence of MSE currently exists and potential
indirect evidence have largely been inconclusive. Space-based studies
have evoked MSE to explain recurring frequencies of both magnetopause
oscillations \cite{plaschke09a,plaschke09b} and narrowband ULF waves
excited by upstream jets \cite{archer13c}, however other mechanisms
could not unambiguously be ruled out and this intepretation of the
results appears inconsistent with later MSE modelling \cite{archer15}.
Multi-instrument ground-based searches in the vicinity of the open-closed
magnetic field line boundary suggest MSE do not occur \cite{pilipenko17,pilipenko18}.
While idealised theoretical treatments of plasmapause surface waves
suggest MSE might be little affected by the ionosphere and thus observable
in ground-based data \cite{kivelson88}, applications of theory specifically
to MSE are currently lacking though and thus it is unclear exactly
what their ground-signatures should be.

One reason perhaps why MSE, if it exists, may not have yet been observed
is that impulsive drivers tend to recur on short time scales and/or
are typically embedded within high levels of turbulence \cite{villante16,plaschke18}.
These perhaps disrupt MSE or result in complicated superpositions
with various other modes of ULF wave. Evidence for other MHD eigenmodes
has relied on multipoint and polarisation observations, comparing
these with theory and simulations \cite{waters02,hartinger13,takahashi15}.
Therefore, multipoint observations of the magnetopause and magnetospheric
response to an isolated impulsive driver may be the ideal scenario
for unambiguous direct evidence of MSE.

Here we present observations at Earth's magnetosphere of an event
which adhered to this strict combination of spacecraft configuration
and driving conditions.  We show that a rare isolated antisunward
plasma jet impinged upon the magnetopause resulting in boundary oscillations
and magnetospheric ULF waves. While the driving jet was impulsive
and broadband, the response was narrowband at well-defined frequencies.
By carefully comparing the observations with the expectations of numerous
possible mechanisms, we show that the response to the jet can only
be explained by the magnetopause surface eigenmode. We therefore present
unambiguous direct observations of this eigenmode, which should exhibit
global effects upon Earth's magnetosphere.

\section*{Results}

\subsection*{Overview}

Observations are taken from the THEMIS mission on 7 August 2007 between
22:10\textendash 22:50~UT, a previously reported interval \cite{dmitriev15,hietala18}.
The spacecraft were ideally arranged in a string-of-pearls configuration
close to the magnetopause in the mid\textendash late morning sector
and $<3^{\circ}$ northwards of the magnetic equatorial plane, as
depicted in Figure~\ref{fig:orbit-timeseries}a-b. Subsequent panels in Figure~\ref{fig:orbit-timeseries}
show time-series observations in the magnetosheath (panels c-d), at the magnetopause (panels e-g),
and within the magnetosphere (panels h-i). The dynamic spectra corresponding to these observations are shown in Figure~\ref{fig:wavelet-phase}a-f.

\subsection*{Magnetosheath Observations}

THB was predominantly located in the region immediately upstream of
the boundary, the magnetosheath, as evidenced by the dominance of
the thermal pressure $P_{\mathrm{th}}$ (red) over the magnetic pressure
$P_{\mathrm{B}}$ (blue) in Figure~\ref{fig:orbit-timeseries}d. At around 22:25~UT, following
an outbound magnetopause crossing, THB observed an antisunward magnetosheath
jet \cite{plaschke18} lasting $\sim100\,\mathrm{s}$ with peak ion
velocity $\sim390\,\mathrm{km}\,\mathrm{s}^{-1}$ directed approximately
along the Sun-Earth line (panels~a-c). An increase in the antisunward
dynamic pressure $P_{\mathrm{dyn},x}$ and thus also the total pressure
acting on the magnetopause $P_{\mathrm{tot},x}=P_{\mathrm{B}}+P_{\mathrm{th}}+P_{\mathrm{dyn},x}$
was associated with the jet (panel~d). Unlike many magnetosheath
jets this structure was isolated with no other significant pressure
variations observed for tens of minutes afterwards \cite{plaschke18}. The solar wind dynamic pressure was steady during this interval (grey line in panel d), with speed (average and spread)
of $609\pm10\,\mathrm{km}\,\mathrm{s}^{-1}$ and density of $2.7\pm0.1\,\mathrm{cm}^{-3}$.
Time-frequency analysis (see Methods) revealed the jet was impulsive
and broadband - power enhancements in the total pressure were contained
within the jet's cone of influence with no statistically significant
peaks at discrete frequencies (Figure~\ref{fig:wavelet-phase}a).

\subsection*{Magnetopause Observations}

The magnetopause passed over four of the spacecraft (THB-E) several
times. Examples of such crossings are shown in Figure~\ref{fig:orbit-timeseries}e-f
for THC, with all crossings indicated as the coloured squares in panel~g
by geocentric radial distance along with the inferred magnetopause
position at all times estimated through interpolation (see Methods).At
least two large-amplitude ($\gtrsim0.4\,\mathrm{R_{E}}$) inward oscillations
of the boundary followed the jet. The first oscillation was largest,
being observed by all four spacecraft, whereas the amplitude had already
decreased by the second oscillation. The wavelet transform of the
interpolated magnetopause position (Figure~\ref{fig:wavelet-phase}b)
shows a narrowband enhancement in power with mean peak frequency $1.8\,\mathrm{mHz}$.

Projections of the normals to the magnetopause, arrived at using the
cross product technique described in the Methods section, form a fan
azimuthally as shown in Figure~\ref{fig:orbit-timeseries}a-b. However,
there was no systematic separation in direction of inbound (purple)
and outbound (orange) normals. Using these normals, timing analysis
was performed (described in Methods) for each inward/outward motion
of the boundary. During the first inward motion of the magnetopause,
concurrent with the jet, the average boundary velocity along the normal
and its spread were $-238\pm76\,\mathrm{km}\,\mathrm{s}^{-1}$ and
showed signs of acceleration with higher velocities resulting when
using later crossings. This magnetopause motion is consistent with
the antisunward ion velocities of the observed magnetosheath jet (Figure~\ref{fig:orbit-timeseries}c).
Therefore, this initial magnetopause motion was a result of the jet's
impulsive enhancement in the total pressure acting on the boundary.
For the subsequent magnetopause motions, the speeds were similar to
one another at $24\pm10\,\mathrm{km}\,\mathrm{s}^{-1}$, consistent
with the $27\,\mathrm{km}\,\mathrm{s}^{-1}$ peak velocities expected
for $0.4\,\mathrm{R_{E}}$ sinusoidal oscillations of the boundary
at $1.8\,\mathrm{mHz}$. Decomposing the boundary velocities into
components normal and transverse to the undisturbed magnetopause (see
Methods) showed that there was little transverse motion ($8\pm8\,\mathrm{km}\,\mathrm{s}^{-1}$).
Indeed, the azimuthal component was consistent with zero ($-1\pm12\,\mathrm{km}\,\mathrm{s}^{-1}$).
No systematic differences between inbound and outbound crossings were
present within these results.

At 22:22:30~UT, before the magnetosheath jet, a $\sim250\,\mathrm{km}\,\mathrm{s}^{-1}$
reconnection outflow \cite{hietala18} was observed during a magnetopause
crossing (Figure~\ref{fig:orbit-timeseries}c), however, no further
clear evidence of local reconnection occurred during subsequent crossings,
likely because the observed magnetic shears were low (mean and spread
were $34\pm22^{\circ}$).

\subsection*{Magnetosphere Observations}

\begin{figure}
\begin{centering}
\includegraphics[width=1\columnwidth]{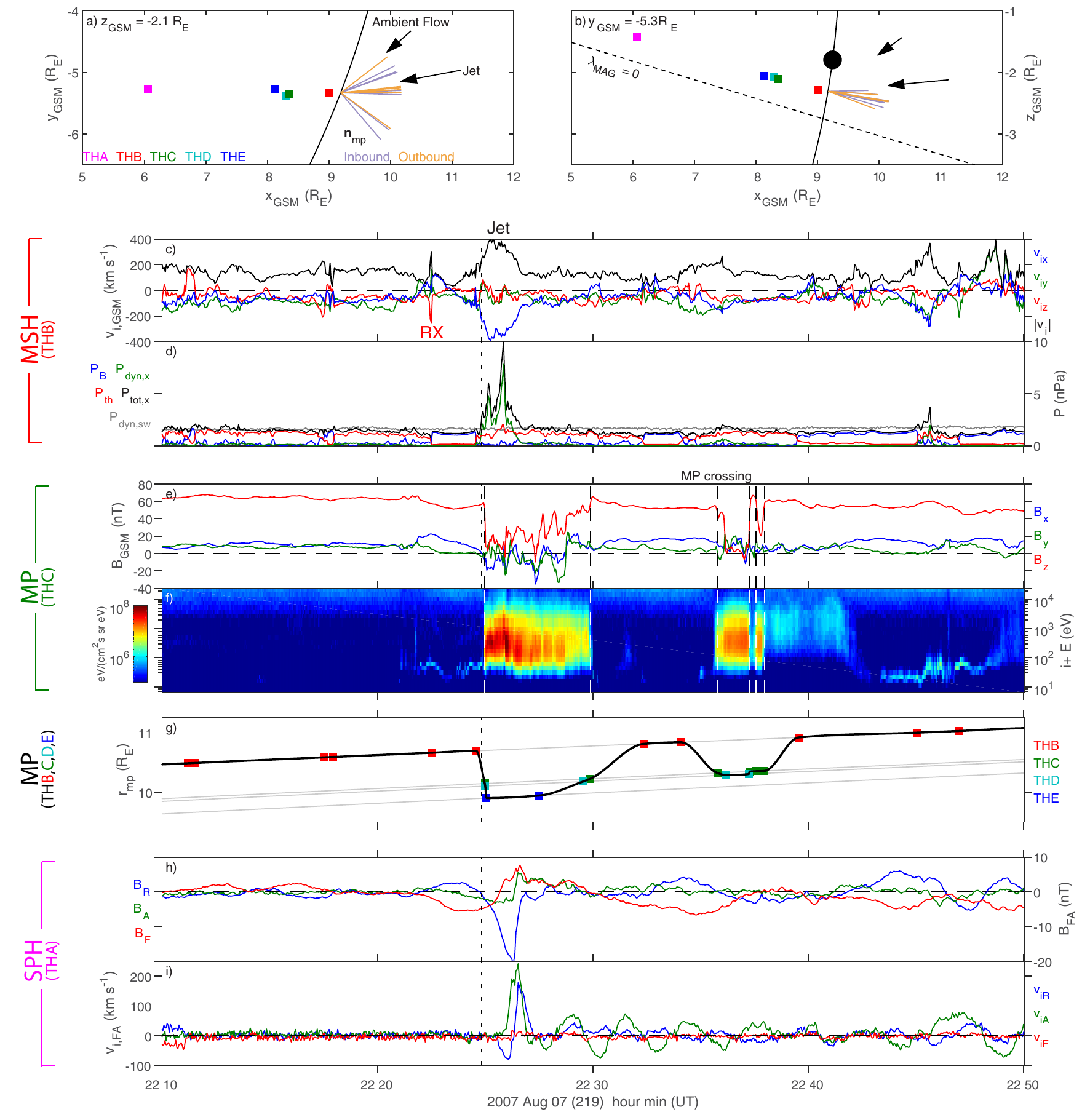}
\par\end{centering}
\caption{\textbf{THEMIS spacecraft locations and observations} (a-b) Projections
of the THEMIS spacecraft positions in the $z_{GSM}=-2.1\,\mathrm{R_{E}}$
(a) and $y_{GSM}=-5.3\,\mathrm{R_{E}}$ (b) planes. Lines indicate
the model magnetopause \cite{shue98} (solid) and magnetic equator
(dotted). Observed magnetopause normals from inbound (purple) and
outbound (orange) crossings are also shown. The black dot marks the
expected location of MSE phase midpoint \cite{archer15}. (c) Ion
velocity at THB in GSM (x, y, z as blue, green, red) and its magnitude
(black). A reconnection exhaust is indicated by RX. (d) Magnetic (blue),
thermal (red), antisunward dynamic (green) and total antisunward (black)
pressures at THB along with lagged solar wind dynamic pressure observations
by Wind (grey). (e) Magnetic field at THC in GSM (colours as before).
(f) Omnidirectional ion energy flux at THC. (g) THEMIS magnetopause
crossings as a function of geocentric radial distance (coloured squares)
with the interpolated magnetopause location shown in black. (h) Magnetic
field perturbations at THA in field-aligned (FA) coordinates (radial,
azimuthal, field-aligned as blue, green, red). (i) Ion velocity perturbations
at THA in FA co-ordinates (colours as before). Vertical dotted lines
indicate times of the magnetosheath jet whereas dashed lines indicate
magnetopause crossings.\label{fig:orbit-timeseries}}
\end{figure}

\begin{figure}
\begin{centering}
\includegraphics[width=1\columnwidth]{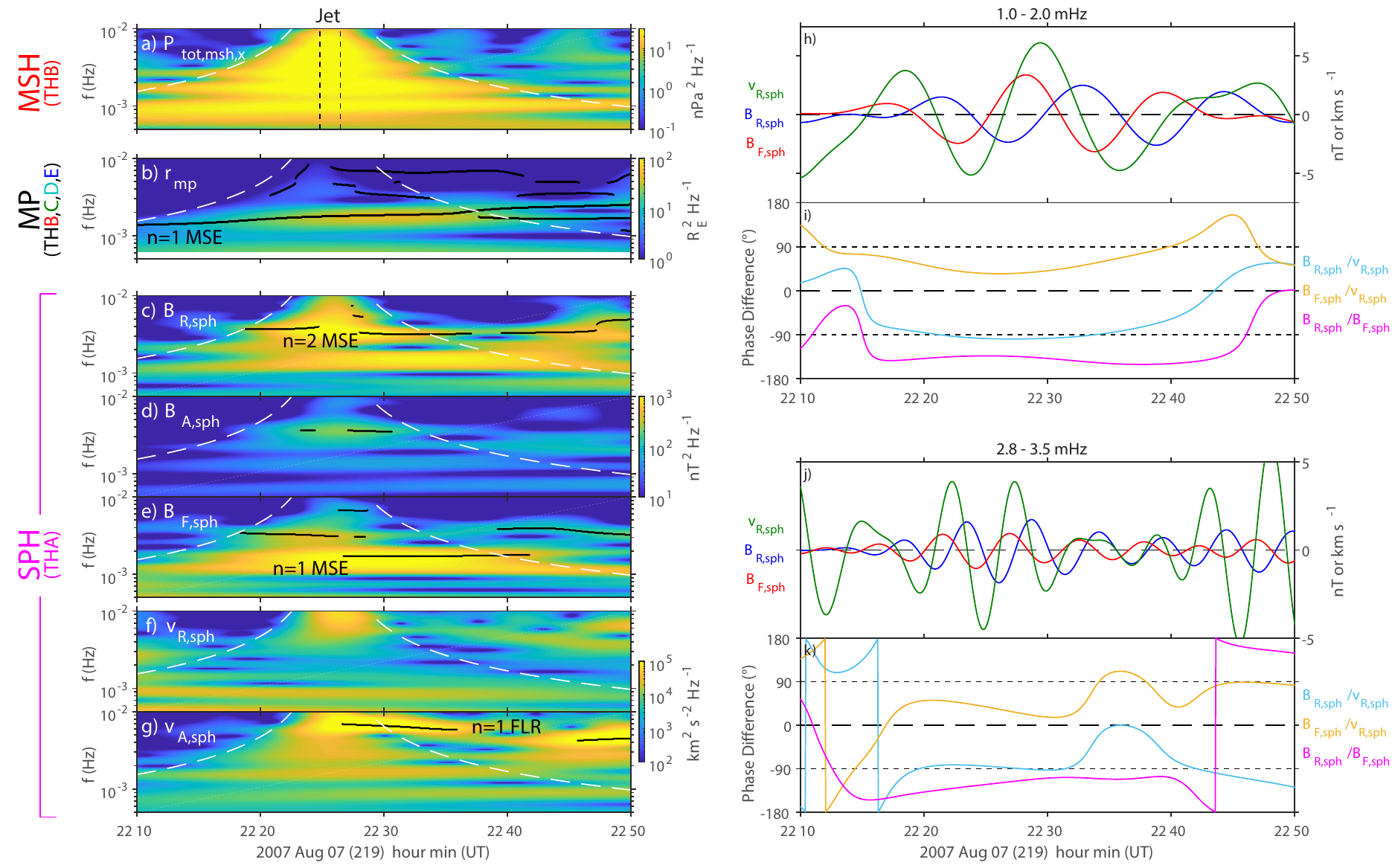}
\par\end{centering}
\caption{\textbf{Observed dynamic spectra and phase relationships }(a-g) Wavelet
dynamic power spectra of the magnetosheath total antisunward pressure
(a), magnetopause location (b), magnetospheric radial (c), azimuthal
(d) and field-aligned (e) magnetic field perturbations, and magnetospheric
radial (f) and azimuthal (g) ion velocity perturbations. Statistically
significant peaks are indicated by black lines. The times of the magnetosheath
jet (black dotted) and its cone of influence (white dashed) are also
shown. (h-k) Wavelet band-pass filtered perturbations of the magnetospheric
radial velocity (green) and radial (blue) and field-aligned (red)
magnetic field pertubations at THA (h,j) along with their cross phases
(i,k) where cyan is the difference between radial magnetic field and
radial velocity, yellow is between the field-aligned magnetic field
and radial velocity, and magenta is between the radial and field-aligned
magnetic fields.\label{fig:wavelet-phase}}
\end{figure}

The magnetopause did not pass over THA and thus it provided uninterrupted
observations of the outer magnetosphere in the vicinity of the magnetopause.
The magnetic field and ion velocity observations are shown in Figure~\ref{fig:orbit-timeseries}h-i
with corresponding wavelet spectra in Figure~\ref{fig:wavelet-phase}c-g.
An initial large-amplitude transient was observed immediately following
the jet, chiefly in the radial components of the magnetic field $B_{R,\mathrm{sph}}$
and ion velocity $v_{\mathrm{i}R,\mathrm{sph}}$ as well as the azimuthal
ion velocity $v_{\mathrm{i}A,\mathrm{sph}}$. Longer period ULF wave
activity occurred afterwards. The field-aligned magnetic field perturbation
$B_{F,\mathrm{sph}}$ showed a $1.7\,\mathrm{mHz}$ signal (Figure~\ref{fig:wavelet-phase}e),
in approximate antiphase to the magnetopause location (Figure~\ref{fig:orbit-timeseries}g-h).
While the $B_{R,\mathrm{sph}}$ timeseries appeared to exhibit a similar
but opposite signal to $B_{F,\mathrm{sph}}$ (Figure~\ref{fig:orbit-timeseries}h),
this did not satisfy our significance test. $B_{R,\mathrm{sph}}$
did, however, feature significant oscillations peaked at $3.3\,\mathrm{mHz}$
(Figure~\ref{fig:wavelet-phase}c). The $v_{\mathrm{i}R,\mathrm{sph}}$
timeseries exhibited some small-amplitude complex oscillations on
timescales potentially consistent with those observed in the magnetic
field and boundary location (Figure~\ref{fig:orbit-timeseries}i),
however the wavelet transform revealed no statistically significant
periodicities. A clear $6.7\,\mathrm{mHz}$ signal dominated $v_{\mathrm{i}A,\mathrm{sph}}$
(Figures~\ref{fig:orbit-timeseries}i and \ref{fig:wavelet-phase}g),
a higher frequency than those previously discussed. No appreciable
variations were present in $v_{\mathrm{i}F,\mathrm{sph}}$. Note that
none of the statistically significant signals commenced before the
magnetosheath jet's cone of influence (white dashed lines in Figure~\ref{fig:wavelet-phase}a-g)
and therefore these oscillations did not precede the jet. 

It is surprising that no obvious radial velocity perturbations associated
with the magnetopause motion were present, regardless of whether this
motion was associated with an eigenmode. However, through modelling
(see Methods) we find that the expected $\sim27\,\mathrm{km}\,\mathrm{s}^{-1}$
amplitude velocity oscillations based on the magnetopause motion would
only be detected as $6\,\mathrm{km}\,\mathrm{s}^{-1}$ due to instrumental
effects associated with cold magnetospheric ions and the spacecraft
potential. The amplitude of $1.0\text{\textendash}2.0\,\mathrm{mHz}$
band radial velocity perturbations were in good agreement with this,
as shown in Figure~\ref{fig:wavelet-phase}h.

We investigate the phase relationships between the three signals present
in the THA data (Figure~\ref{fig:wavelet-phase}h-k). Similar coherent
phase relationships were found for the two lower frequency signals
with $B_{R,\mathrm{sph}}$ in quadrature with $v_{\mathrm{i}R,\mathrm{sph}}$
(means and spreads of$-96\pm4^{\circ}$ and $-86\pm4^{\circ}$ for
the $1.0\text{\textendash}2.0\,\mathrm{mHz}$ and $2.8\text{\textendash}3.5\,\mathrm{mHz}$
bands respectively) and some $50^{\circ}$ away from antiphase with
$B_{F,\mathrm{sph}}$ ($-138\pm5^{\circ}$ and $-123\pm8^{\circ}$),
as well as the phase between $B_{F,\mathrm{sph}}$ and $v_{\mathrm{i}R,\mathrm{sph}}$
being consistent with $50^{\circ}$ out from quadrature ($-42\pm8^{\circ}$
and $-37\pm12^{\circ}$). In the $4.9\text{\textendash}8.6\,\mathrm{mHz}$
band $v_{\mathrm{i}A,\mathrm{sph}}$ led $B_{A,\mathrm{sph}}$ by
$82\pm6^{\circ}$, likely indicating a toroidal field line resonance
(FLR, a standing Alfv\'{e}n wave) \cite{takahashi15}.

\subsection*{Solar Wind Observations}

\begin{figure}
\begin{centering}
\includegraphics[width=1\columnwidth]{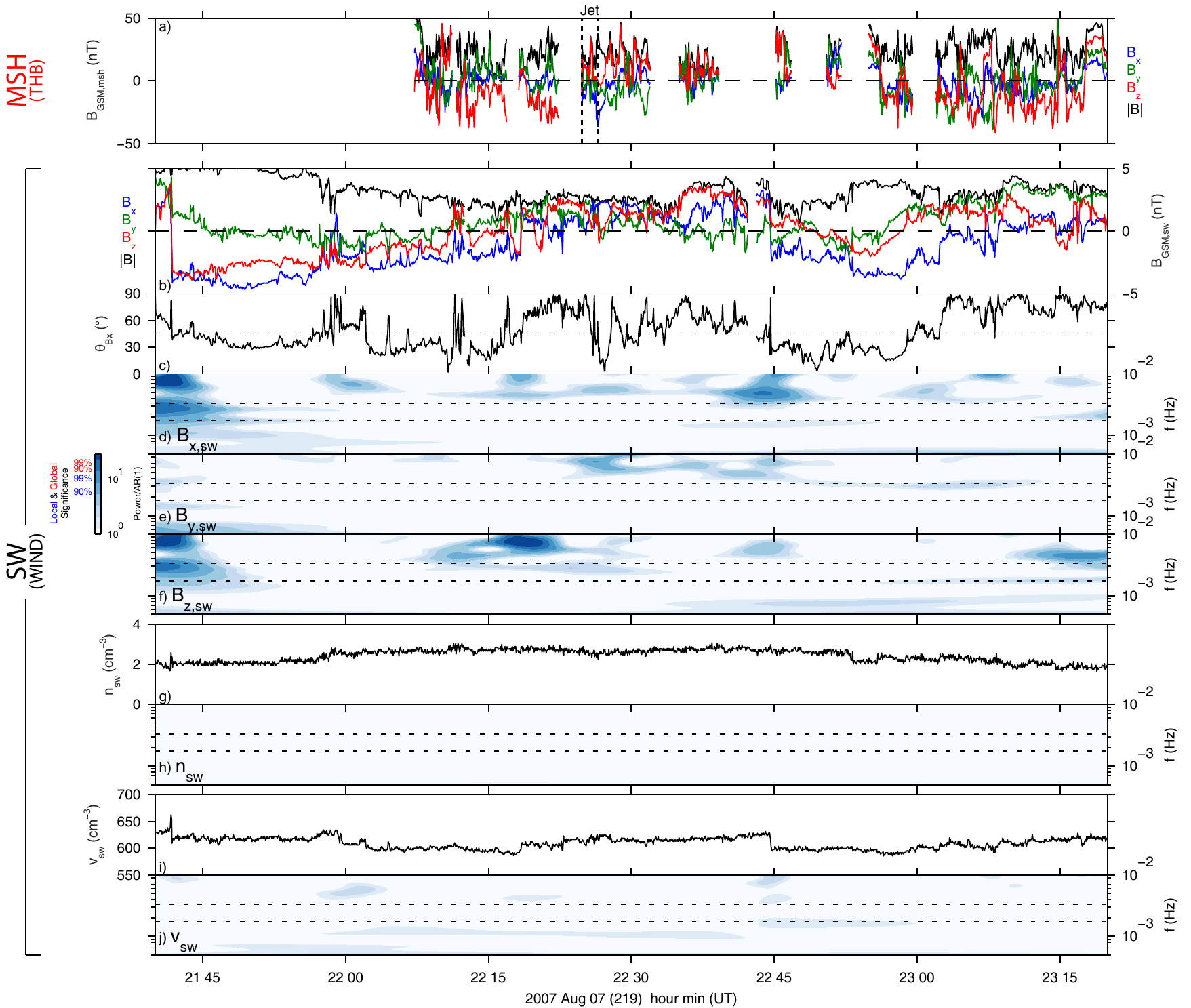}
\par\end{centering}
\caption{\textbf{Upstream solar wind observations }(a) Magnetosheath magnetic
field at THB in GSM components (x, y, z as blue, green, red) and magnitude
(black). Observations within the magnetosphere have been removed for
clarity. The times of the magnetosheath jet are shown by vertical
black dotted lines. (b-j) Lagged Wind observations of the pristine
solar wind (b) magnetic field GSM components (x, y, z as blue, green,
red) and magnitude (black), (c) cone angle, (g) density, and (i) speed.
The significance of their respective wavelet spectra are also shown
(d,e,f,h,j), where the power has been divided by an autoregressive
noise model. Dotted horizontal lines depict frequencies of $1.7\text{\textendash}1.8$
and $3.3\,\mathrm{mHz}$.\label{fig:sw}}
\end{figure}

While the solar wind dynamic pressure was steady throughout this period,
a number of fluctuations in the interplanetary magnetic field (IMF)
were present, shown in Figure~\ref{fig:sw}b, particularly with several
sign reversals in $B_{z,\mathrm{sw}}$. Many of these fluctuations
were transmitted to the magnetosheath and observed by THB, as shown
in panel~a where observations within the magnetosphere have been
removed for clarity. It can be seen that some of these sign reversals
in fact precede the magnetosheath jet. While the magnetosheath magnetic
field observations were sparse and rather turbulent, there is an apparent
near one-to-one correspondence between the sign reversals in the solar
wind and magnetosheath observations during the period of interest
(see Methods for details of the lagging procedure). Nonetheless, we
present an additional $30\,\mathrm{min}$ of solar wind data either
side of the interval to allow for possible errors.

The magnetosheath jet occurred around the time of a magnetic field
rotation which changed the IMF cone angle (the acute angle between
the IMF and the Sun-Earth line) and thus the character of the bow
shock upstream of the THEMIS spacecraft. When the cone angle is below
$\sim45^{\circ}$ the subsolar bow shock is quasi-parallel, whereby
suprathermal particles can escape far upstream leading to various
nonlinear kinetic processes \cite{eastwood05}. This results in a
much more complicated shock region and turbulent magnetosheath downstream,
with various transient phenomena that can impinge upon the magnetopause
e.g. magnetopause surface oscillations occur more frequenctly under
low cone angle conditions likely because of such transients \cite{plaschke09b}.
Magnetosheath jets are just one example, with some of the strongest
jets being caused by changes in the IMF orientation from quasi-perpendicular
to quasi-parallel conditions \cite{archer12}, as appeared to be the
case during this event. Following this short period of low cone angle
IMF, the shock conditions were oblique or quasi-perpendicular for
most of the rest of the interval.

The variations present in the upstream solar wind did not appear to
be periodic. The statistical significance of the wavelet power compared
to autoregressive noise is shown for the three components of the IMF
(Figure~\ref{fig:sw}d-f) as well as for the solar wind density (Figure~\ref{fig:sw}h)
and speed (Figure~\ref{fig:sw}j). Throughout the extended interval
presented, there were very few enhancements in wavelet power for any
of the quantities considered that were even locally significant (let
alone the more strict global significance we have imposed on the THEMIS
observations). Crucially, there were no significant enhancements peaked
at (or near) either $1.7\text{\textendash}1.8$ or $3.3\,\mathrm{mHz}$
frequencies (indicated by the horizontal dotted lines).

Given that the aperiodic IMF variations were present before the jet
but the magnetopause motions and magnetospheric ULF waves all occurred
directly following it, we conclude that the magnetosheath jet was
indeed the driver of the narrowband signals observed by THEMIS.

\subsection*{Eigenfrequency estimates}

\begin{figure}
\begin{centering}
\includegraphics{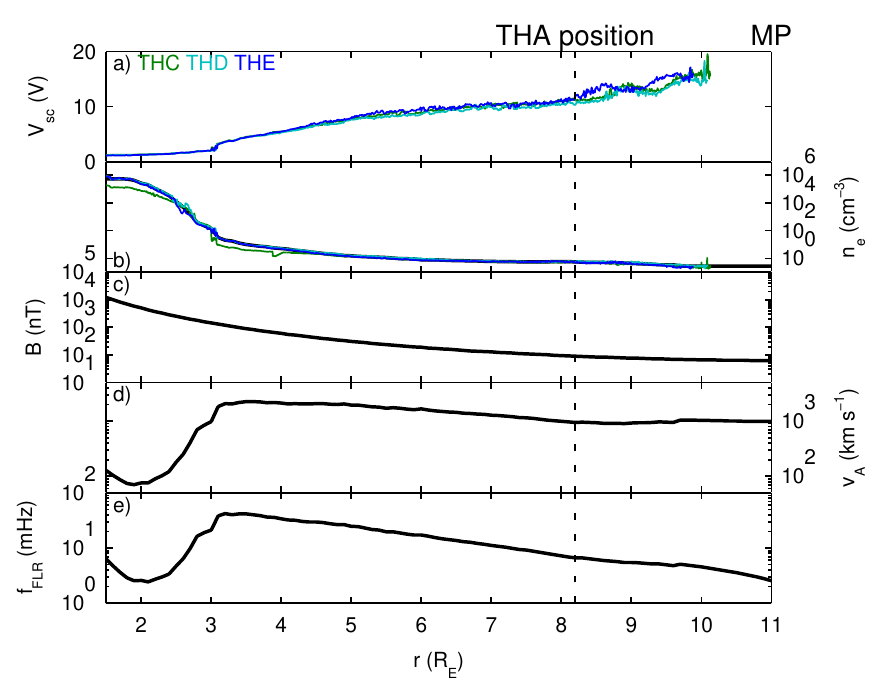}
\par\end{centering}
\caption{\textbf{Magnetospheric radial profiles} (a) spacecraft potentials,
(b) potential inferred electron densities, (c) T96 magnetic field,
(d) Alfv\'{e}n speed, (e) fundamental Field Line Resonance (FLR)
frequency. THA's location is indicated as the dotted line.\label{fig:profiles}}
\end{figure}

To aid in our interpretation of the observed signals, we compare their
frequencies with estimates of various resonant ULF wave modes applied
to this event using the WKB method. From an existing database of numerical
calculations within representative models \cite{archer15} the $n=1$
MSE is expected at $1.4\,\mathrm{mHz}$ during this interval, with
its antinode located at the black circle in Figure~\ref{fig:orbit-timeseries}b.
Spacecraft potential observations from THD and THE were used to arrive
at the radial profile of the electron density \cite{mcfadden08b}
shown in Figure~\ref{fig:profiles}b (black). See Methods for details.
We combine the resulting density profile with a T96 magnetospheric
magnetic field model \cite{t95,t96} using hourly averaged upstream
conditions, an average ion density of $6.8\,\mathrm{amu}\,\mathrm{cm}^{-3}$
\cite{lee14}, and assuming a power law for the density distribution
along the field line using exponent 2 \cite{denton02}. Fundamental
field line resonance (FLR) frequencies are then given at each radial
distance by 
\begin{equation}
f_{\mathrm{FLR}}=\left(2\int\frac{dF}{v_{\mathrm{A}}}\right)^{-1}
\end{equation}
where $v_{\mathrm{A}}$ is the local Alfv\'{e}n speed and the integration
occurs between the two footpoints of each field line, with the results
shown in Figure~\ref{fig:profiles}e. At THA's location this is estimated
to be $6.7\,\mathrm{mHz}$ (panel~e) in excellent agreement with
the observed signal in $v_{\mathrm{i}A,\mathrm{sph}}$, hence the
observed frequency, polarisation and relative amplitudes point towards
this signal being an $n=1$ toroidal FLR.

Fast-mode resonances (FMRs), also known as cavity or waveguide modes,
are radially standing fast-mode waves between boundaries and/or turning
points \cite{kivelson84,kivelson85}. In the outer magnetosphere,
the lowest frequency FMRs are quarter wavelength modes resulting from
over-reflection of fast-mode waves. It is thought that these may occur
for magnetosheath flow speeds $\gtrsim500\,\mathrm{km}\,\mathrm{s}^{-1}$
\cite{mann99}. However, at the local times of the observations this
was not satisfied for either the ambient or the jet's flow speeds.
Nonetheless, we still estimate the lowest possible FMR frequency given
by
\begin{equation}
f_{\mathrm{FMR}}=\left(4\int_{r_{\mathrm{ib}}}^{r_{\mathrm{mp}}}\frac{dR}{v_{\mathrm{A}}}\right)^{-1}
\end{equation}
This corresponds to a fast-mode wave propagating (assuming low plasma
beta) purely in the $\pm R$ direction forming a quarter wavelength
mode between the magnetopause $r_{\mathrm{mp}}$ and an inner boundary
at the Alfv\'{e}n speed local maximum $r_{\mathrm{ib}}$ (at $r=3.2\,\mathrm{R_{E}}$)
\cite{archer17}. From the Alfv\'{e}n speed profile for this event
we calculate this to be $6.3\,\mathrm{mHz}$, clearly much higher
than the two remaining signals which were observed.

\subsection*{Ground Magnetometer Observations}

\begin{figure}
\begin{centering}
\includegraphics{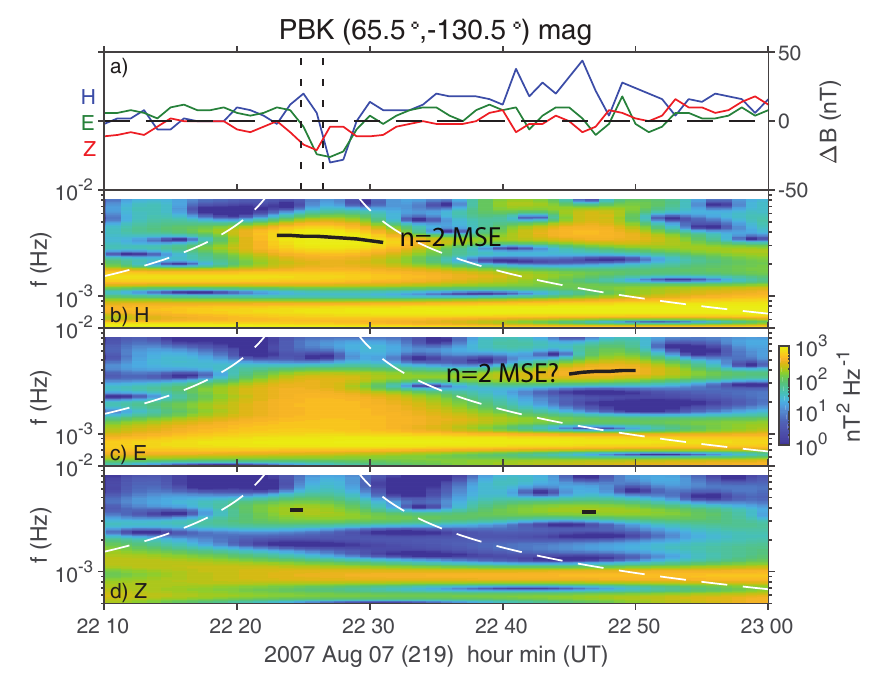}
\par\end{centering}
\caption{\textbf{Conjugate ground magnetometer observations at Pebek} (a) magnetic
deflections in geomagnetic co-ordinates (H, E, Z as blue, green, red).
(b-d) Wavelet dynamic power spectra of the H (b), E (c) and Z (d)
components in the same format as Figure~\ref{fig:wavelet-phase}.\label{fig:gmag}}
\end{figure}

Unfortunately, there was very poor ground magnetometer station coverage
near the spacecrafts\textquoteright{} footpoints with only one station
available, Pebek (PBK; see Methods for selection criteria). This station
was nearly conjugate with THA, whose footpoint was at (66.3\textdegree ,
-132.0\textdegree ) geomagnetic latitude and longitude respectively.
The observations are shown in Figure~\ref{fig:gmag}.

A transient, similar to that at THA immediately following the jet,
was observed in the H and E components. Its timing was consistent
with the $\sim40\,\mathrm{s}$ Alfv\'{e}n travel time from the equatorial
magnetosphere to the ground. Similar to the THA observations, following
this transient other oscillations also occurred. Time-frequency analysis
identified several statistically significant signals. In the H component
this peaked at $3.5\pm0.2\,\mathrm{mHz}$ and was contained within
the jet's cone of influence. A later signal following the jet's cone
of influence was present in the E component at $3.9\pm0.1\,\mathrm{mHz}$.
The former was likely the ground signature of the $3.3\,\mathrm{mHz}$
signal observed by THA, however it is not entirely clear if this is
also the case with the latter and if so why a change in polarisation
occurred. Both these signals in the ground data had corresponding
signatures in the Z componont, though these were weak and very short
lived (only 2 datapoints for each were statistically significant).
While a power enhancement consistent with the $1.7\text{\textendash}1.8\,\mathrm{mHz}$
signal could be seen in the H component, this did not satisfy our
significance test. Finally, the $6.7\,\mathrm{mHz}$ toroidal FLR
at THA might be expected in the H component on the ground due to the
approximate 90\textdegree{} rotation of Alfv\'{e}n waves by the ionosphere
\cite{hughes76a}. However, its frequency was not well resolved by
the coarse data being only 20\% lower than the Nyquist frequency.
Nonetheless, the FLR was likely the cause of the triangular wave-like
oscillations present in this component following the initial transient.

The poor coverage and low resolution of the ground magnetometer data
mean it is insufficient in providing additional evidence towards the
physical mechanism behind the THEMIS observations.

\section*{Discussion}

\begin{figure}
\begin{centering}
\includegraphics[width=1\columnwidth]{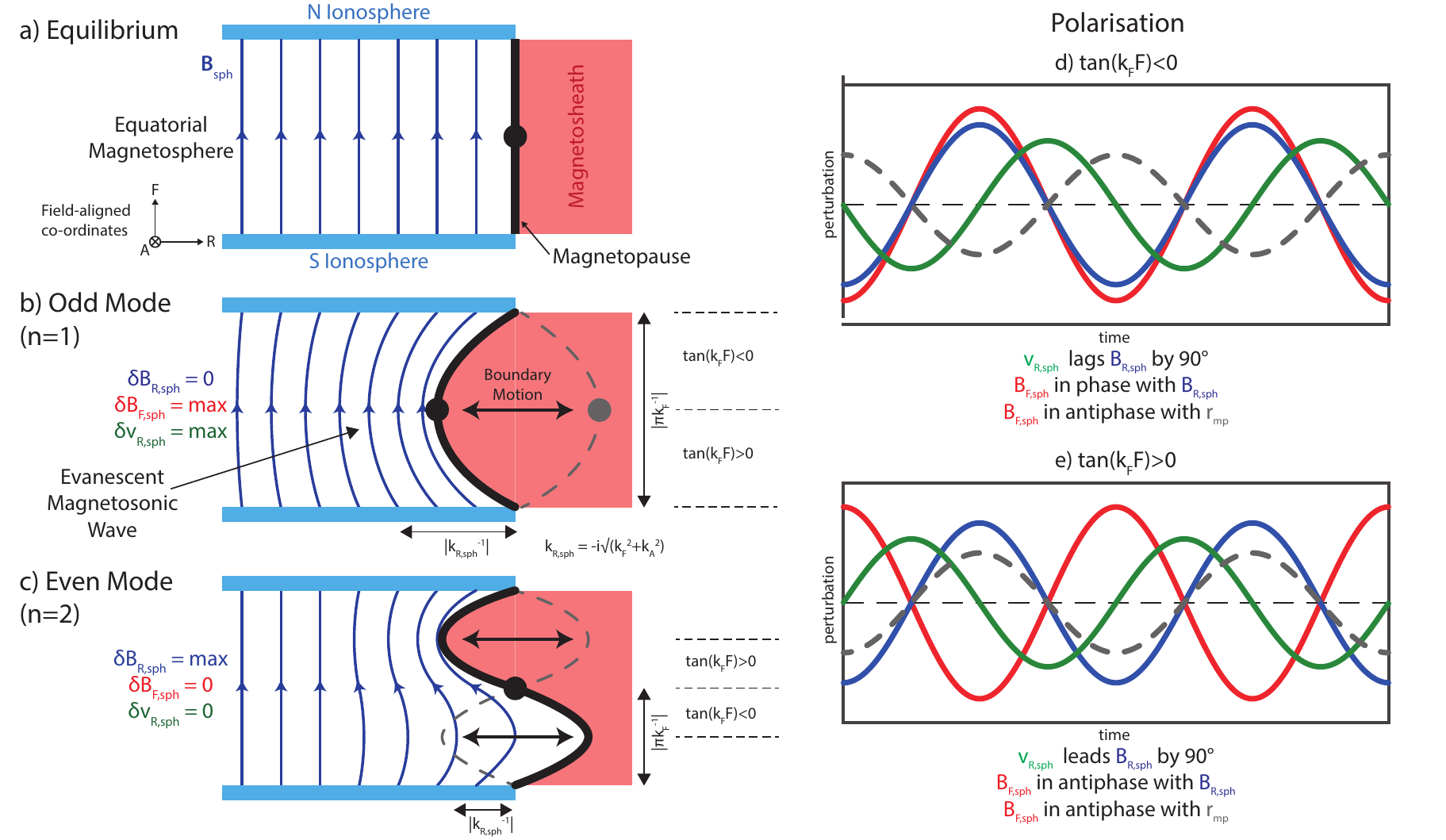}
\par\end{centering}
\caption{\textbf{Schematic of the magnetopause surface eigenmode in a box model
} a) Box model equilibrium featuring the magnetopause
(black) separating the magnetosheath (red) and magnetosphere (dark
blue arrows depict the geomagnetic field bounded by the the northern
and southern ionospheres coloured light blue). The directions of the
field-aligned coordinate system in this model are also shown where
$R$ is radial, $A$ azimuthal and $F$ field-aligned. Subsequent
panels depict $n=1$ (b) and $n=2$ (c) MSE. The midpoint of the phase is
indicated as the black dot, which corresponds to the location of the
MSE $n=1$ antinode and $n=2$ node. Expected MSE polarisations in
different regions of the magnetosphere for the magnetopause standoff
distance (grey dashed), radial velocity (green), radial (blue) and
field-aligned (red) magnetic field components are shown on the right (d-e).\label{fig:cartoon}}
\end{figure}

We have presented THEMIS observations of the magnetopause and magnetospheric
response to an isolated, impulsive antisunward magnetosheath jet.
The $\sim100\,\mathrm{s}$ duration jet triggered narrowband oscillations
of both the magnetopause at $1.8\,\mathrm{mHz}$ and magnetospheric
ULF waves with peak frequencies of $1.7$, $3.3$, and $6.7\,\mathrm{mHz}$.
We now compare the observations with several possible interpretations.
\begin{enumerate}
\item Direct Driving. The solar wind dynamic pressure was steady throughout
this interval and while there were variations present in the IMF,
these were aperiodic. The magnetosheath jet's total pressure was broadband
and impulsive and it has been established from the magnetopause motion
and the start of the wave activity that the jet triggered the observed
signals. Since no significant narrowband oscillations at (or near)
these frequencies were present upstream in either the solar wind or
magnetosheath, we conclude that the observed response cannot have
been directly driven.
\item Propagating Alfv\'{e}n or Fast-Mode Waves. The associated perturbations
in $\boldsymbol{v}_{\mathrm{sph}}$ and $\mathbf{B}_{\mathrm{sph}}$
should either be in-phase or antiphase, unlike the observations. Furthermore,
neither of these modes can explain the magnetopause motion nor the
origin of the narrowband signals given the broadband driver.
\item Propagating Magnetopause Surface Waves. From linear analysis, the
magnetospheric signature of a propagating surface wave should exhibit
an in-phase/antiphase relationship between $\boldsymbol{v}_{\mathrm{sph}}$
and $\mathbf{B}_{\mathrm{sph}}$ as well as quadrature between $B_{R,\mathrm{sph}}$
and $B_{F,\mathrm{sph}}$ \cite{plaschke11}, neither of which was
observed in this event. Furthermore, while the fanning out of magnetopause
normals azimuthally is consistent with travelling surface waves, perhaps
due to the Kelvin-Helmholtz instability, the lack of a difference
between inbound and outbound crossings is not \cite{lepping79} assuming
linear waves. There is no evidence from the multipoint interpolated
magnetopause position for non-linear overturning surface waves, pointing
instead to a simple wave pattern. Crucially, timing analysis of the
boundary (unaffected by assumptions of linearity) revealed the motions
were largely directed along the normal to the undisturbed magnetopause,
with azimuthal velocities consistent with zero i.e. no transverse
propagation.
\item Field Line Resonance. We have already concluded that the $6.7\,\mathrm{mHz}$
signal corresponded to a fundamental toroidal FLR at THA because of
the observed polarisation and excellent agreement with the estimated
frequency of this mode. The $v_{\mathrm{i}R,\mathrm{sph}}\text{\textendash}B_{R,\mathrm{sph}}$
phase relationships for the $1.7\text{\textendash}1.8$ and $3.3\,\mathrm{mHz}$
signals could be consistent with poloidal FLRs \cite{takahashi15}.
The poloidal mode is known to have slightly lower natural frequencies
than the toroidal, however, these differences are typically no more
than 15\textendash 30\% \cite{rankin06}. Therefore, given that the
$n=1$ toroidal FLR frequency at THA was $6.7\,\mathrm{mHz}$ during
this event, the much lower frequencies of $1.7\text{\textendash}1.8$
and $3.3\,\mathrm{mHz}$ cannot be explained as poloidal FLRs. Additionally,
magnetopause motion is not expected to result from an FLR located
several $\mathrm{R_{E}}$ Earthward of the boundary.
\item Fast-Mode Resonance. Observational signatures of radially standing
fast-mode waves require $\pm90^{\circ}$ phase differences between
$v_{\mathrm{i}R,\mathrm{sph}}$, equivalent to the azimuthal electric
field via $\mathrm{\boldsymbol{E}}=-\boldsymbol{\mathrm{v}}\times\boldsymbol{\mathrm{B}}$,
and $B_{F,\mathrm{sph}}$ \cite{waters02,hartinger13}, which were
not observed. Exceptions to this perhaps occur in cases of exceptionally
leaky or over-reflecting boundaries, however this would not be the
case at the local times of the observations due to the moderate flow
speeds present \cite{mann99}. The large amplitude magnetopause motions
with near-zero azimuthal phase velocities are also inconsistent with
a fast-mode resonance interpretation. Finally, we estimate that during
this event cavity/waveguide modes of any type cannot explain frequencies
below $6.3\,\mathrm{mHz}$. The difference between this estimate and
the observed lower frequency signals are much larger than the expected
errors ($\sim3\%$ \cite{rickard95}).
\item Pulsed Reconnection. While a reconnection outflow was seen before
the magnetosheath jet, no clear signatures of local magnetopause reconnection
were observed subsequently throughout the event.
\item Magnetopause Surface Eigenmode. The $1.4\,\mathrm{mHz}$ estimated
fundamental MSE frequency during this period agrees with the observed
$1.7\text{\textendash}1.8\,\mathrm{mHz}$ signal within errors \cite{hartinger15,archer15},
with the $3.3\,\mathrm{mHz}$ oscillation perhaps being the second
harmonic. As depicted in Figure~\ref{fig:cartoon}b, equatorial observations
of an $n=1$ mode should show strong signals in the motion of the
magnetopause as well as $v_{\mathrm{i}R,\mathrm{sph}}$ and $B_{F,\mathrm{sph}}$,
whereas an $n=2$ mode should dominate simply in $B_{R,\mathrm{sph}}$ (panel c).
These are all in agreement with the statistically significant peaks
in the wavelet spectra, after the instrumental effects on the ion
velocity due to the spacecraft potential were modelled and taken into
account. The similarity in observed magnetopause normals for inbound
and outbound crossings as well as an azimuthal boundary velocity consistent
with zero are both expected for a standing surface wave. The phase
relationships between the quantities for both signals were in good
agreement with theoretical expectations of MSE \cite{plaschke11}
in the regions $\tan k_{F}F>0$ as depicted in Figure~\ref{fig:cartoon}e
when also taking into account the reported $50^{\circ}$ phase shift
of $B_{F,\mathrm{sph}}$ in global MHD simulations of MSE \cite{hartinger15}.
Given the spacecraft were just southward of the expected MSE phase
midpoint (Figure~\ref{fig:orbit-timeseries}b) this is exactly the
polarisation expected for the fundamental. In contrast, the second
harmonic should see the phase relations for $\tan k_{F}F<0$ in this
region. While in the WKB approximation the $n=1$ antinode and $n=2$
node coincide, this may not be the case in the full solution which
could exhibit anharmonicity as is the case with FLRs \cite{denton02}.
\end{enumerate}
We therefore conclude that THEMIS observed both the $n=1$ and $n=2$
MSEs as the $1.7\text{\textendash}1.8$ and $3.3\,\mathrm{mHz}$ signals
respectively, providing unambiguous direct observations of this eigenmode
made possible only due to the fortuitous multispacecraft configuration
during a rare isolated impulsive magnetosheath jet. MSE constitute
a natural response of the dayside magnetopause, with these observations
at last confirming that plasma boundaries can trap surface wave energy
forming an eigenmode. Magnetopause dynamics in general have wide ranging
effects throughout the entire magnetospheric system and MSE should,
at the very least, act as a global source of magnetospheric ULF waves
that can drive radiation belt / auroral interactions and ionospheric
Joule dissipation.

It remains to be seen how often MSE occur. Future work could search
the large statistical databases of magnetosheath jets for other potential
events (satisfying the strict observational criteria presented in
this paper) to provide further direct evidence. Other impulsive drivers
could also be considered including interplanetary shocks and solar
wind pressure pulses. However, since MSE are difficult to observe
directly, remote sensing methods should be developed. The polarisations
of magnetospheric ULF waves from spacecraft observations, as presented
in this paper, may be one such method. However, potentially more useful
would be ground-based signatures from magnetometers and ionospheric
radar due to the wealth of data being produced. Currently, the ground
signatures of MSE are not well understood, having received little
theoretical attention. However, in this paper we show that MSE can
exhibit at least some similar signals to the in-situ spacecraft observations
within conjugate high-latitude ground magnetometer data. Further investigations
using theory, simulations and observations should explore all possible
remote sensing methods such that the occurrence rates and properties
of MSE more generally can be characterised.

\section*{Methods}

\subsection*{Data}

Observations in this paper are taken from the five Time History of
Events and Macroscale Interactions during Substorms (THEMIS) spacecraft
\cite{angelopoulos08} in particular using the Fluxgate Magnetometers
(FGM) \cite{auster08}, Electrostatic Analysers (ESA) \cite{mcfadden08a}
and Electric Field Instruments (EFI) \cite{bonnell08} all at $3\,\mathrm{s}$
resolution. We used the Geocentric Solar Magnetospheric (GSM) coordinate
system for vector measurements from all spacecraft except THA. For
this spacecraft, since we use it to evaluate the magnetospheric ULF
wave response, we define a field-aligned (FA) coordinate system. The
linear trend of each GSM magnetic field component was determined between
21:45\textendash 23:30~UT using iteratively reweighted least squares
with bisquare weighting \cite{huber81,street88}. This trend was used
to define the field-aligned direction $\mathbf{F}$ of the FA system
and was subsequently subtracted from the magnetic field data. The
azimuthal direction $\mathbf{A}$, which nominally pointed eastward,
was given by the cross product of $\mathbf{F}$ with the spacecraft's
geocentric position. Finally the radial direction, predominantly directed
radially outwards from the Earth, was determined by $\mathbf{R}=\mathbf{A}\times\mathbf{F}$.
The equivalent directions of the FA system in the MSE box model are
shown in Figure~\ref{fig:cartoon}.

Solar wind observations at the L1 Lagrange point were taken from the
Wind spacecraft's 3-D Plasma and Energetic Particle Investigation
\cite{lin95} and Magnetic Field Investigation \cite{lepping95} both
at $3\,\mathrm{s}$ resolution. In order for this data to approximately
correspond to the shocked solar wind arriving in the vicinity of the
magnetopause, a constant time lag was applied. First the data was
time lagged by $40\,\mathrm{min}\,27\,\mathrm{s}$, the average amount
given in the OMNI dataset from the Wind spacecraft to the bow shock
nose. An additional $2\,\mathrm{min}$ lag to the magnetopause was
subsequently added, determined by manually matching up sign reversals
in the solar wind magnetic field observations with those in the magnetosheath
at THB (see Figure~\ref{fig:sw}a-b). Using Advanced Composition
Explorer (ACE) solar wind data instead of Wind did not substantially
change any of the subsequent results.

Finally, ground magnetometer data was also used. Ground stations were
chosen by computing the locations of the footpoints of the THEMIS
spacecraft from a T96 model \cite{t95,t96}. Only ground stations
on closed field lines (according to T96) no more than $1\,\mathrm{R_{E}}$
earthward from the observations and within $\pm1\,\mathrm{hr}$ of
magnetic local time were selected. This, unfortunately, resulted in
only one station, Pebek (PBK) in the Russian Arctic. Data from this
station was only available at 60~s resoluion and are presented in
geomagnetic coordinates where the horizontal components H and E point
geomagnetically north and east respectively and Z is the vertical
component. The median was subtracted from each component.

\subsection*{Magnetopause motion}

To track the location and motion of the magnetopause, the innermost
edge of the magnetopause current layer was identified manually from
THEMIS FGM data and piecewise cubic hermite interpolating polynomials
\cite{fritsch80} were used to estimate the radial distance to the
boundary from all crossings (shown as the coloured squares in Figure~\ref{fig:orbit-timeseries}g)
at all times, resulting in the black line. This method was chosen
because it does not suffer from overshooting and anomalous extrema
as much as other spline interpolation methods, thus any resulting
oscillations present would be underestimates. Nonetheless, the crucial
aspects of the results presented, such as the time-frequency analysis,
proved to be largely insensitive to the interpolation method used.

Boundary normals for each magnetopause crossing were also estimated.
This was done by taking the cross product of $30\,\mathrm{s}$ averages
of magnetic field observations either side of each crossing, which
assumes that the magnetopause was a tangential discontinuity \cite{schwartz98}.
This method was used since minimum variance analysis \cite{sonnerup98}
was poorly conditioned throughout the interval (the ratio of intermediate
to minimum eigenvalues was $\sim2$). The normals were insensitive
to the precise averaging period used. Projections of these normals
are shown in Figure~\ref{fig:orbit-timeseries}a-b where we distinguish
between inbound and outbound crossings by colour. Magnetic shear angles
were calculated from the same averaged magnetic field observations.

Finally, two-spacecraft timing analysis was also performed. Using
the ascertained magnetopause normals $\boldsymbol{\mathrm{n}}$, the
velocity of the boundary along the normal is given by

\begin{equation}
v_{n}=\boldsymbol{\mathrm{n}}\cdot\left(\boldsymbol{\mathrm{r}}_{\alpha}-\boldsymbol{\mathrm{r}}_{\beta}\right)/\left(t_{\alpha}-t_{\beta}\right)
\end{equation}
where $\boldsymbol{r}_{\alpha}$ is the position of spacecraft $\alpha$
during the magnetopause crossing at time $t_{\alpha}.$ This assumes
a planar surface with constant speed. For each inward/outward motion
of the magnetopause, the analysis was applied to all spacecraft pairs
using both sets of normals. The multiple THC crossings at around 22:37~UT
were neglected. Taking the average magnetopause normal over all crossings
$\boldsymbol{\mathrm{N}}$ as representative of the undisturbed boundary, each
determined magnetopause velocity can be decomposed into parallel and
perpendicular velocities

\begin{align}
\boldsymbol{\mathrm{v}}_{\Vert}= & v_{n}\left(\boldsymbol{\mathrm{n}}\cdot\boldsymbol{\mathrm{N}}\right)\boldsymbol{\mathrm{N}}\\
\boldsymbol{\mathrm{v}}_{\bot}= & v_{n}\boldsymbol{\mathrm{n}}-v_{n}\left(\boldsymbol{\mathrm{n}}\cdot\boldsymbol{\mathrm{N}}\right)\boldsymbol{\mathrm{N}}
\end{align}
Replacing $\boldsymbol{\mathrm{N}}$ with a normal from a model magnetopause
does not significantly affect the results.

\subsection*{Modelling ESA instrumental effects}

The ESA instrument can only detect ions whose energy overcomes the
spacecraft potential, however the majority of ions in the magnetosphere
are cold \cite{mcfadden08b}. During this interval we find the temperature
of cold ions to be $18\,\mathrm{eV}$ by fitting a Maxwell-Boltzmann
distribution to the population observed in the omnidirectional ion
energy spectrogram at around 22:45~UT (Figure~\ref{fig:orbit-timeseries}f).
While no spacecraft potential observations were available for THA,
those from THC-E suggest a value of $\sim11\,\mathrm{V}$ at THA's
location (Figure~\ref{fig:profiles}a). A sinusoidal oscillation
of the magnetopause $r_{\mathrm{mp}}=C\sin\omega t$ would result
in velocity $v_{\mathrm{i}R,\mathrm{sph}}=C\omega\cos\omega t$ and
using $C=0.4\,\mathrm{R_{E}}$ we find that protons oscillating at$1.8\,\mathrm{mHz}$
would have a peak bulk kinetic energy $\sim4\,\mathrm{eV}$, less
than the assumed spacecraft potential. To estimate the effect on the
data, we take one-dimensional velocity moments of the Boltzmann distribution
corresponding to the cold ions, excluding all energies below the spacecraft
potential. This suggests that the expected velocity oscillations of
$27\,\mathrm{km}\,\mathrm{s}^{-1}$ amplitude would only be detected
as $6\,\mathrm{km}\,\mathrm{s}^{-1}$ by the ESA instrument.

\subsection*{Wavelet transform}

Time-frequency analysis of the data was performed using the Morlet
wavelet transform \cite{torrence98}, with the resulting dynamic power
spectra shown in Figure~\ref{fig:wavelet-phase}a-g. At each time
all peaks between $0.5\text{\textendash}10\,\mathrm{mHz}$ whose power
and prominence were both above the two-tailed global 99\% confidence
interval (using the Bonferonni correction \cite{dunn61}) for an autoregressive
AR(1) noise model were identified, shown as the black lines. The magnetosheath
jet's cone of influence, the region within time-frequency space that
is affected by the jet due to the scale-dependent windowing of the
wavelet transform, are also shown as the white dashed lines. Significant
narrowband signals were investigated by reconstructing a complex-numbered
version of the timeseries from the Morlet wavelet transform across
the bandwidth of each signal only \cite{torrence98}. The real part
of the resulting timeseries is the band-pass filtered data whereas
its phase is used to investigate polarisations. Note that it is not
necessary for both timeseries to exhibit statistically significant
power enhancements in the same region of time-frequency space for
a coherent phase relationship to potentially exist between them within
that region \cite{grinsted04}.

\subsection*{Spacecraft-potential inferred density}

The electron density can be inferred from measurements of a spacecraft's
potential and in this paper we use an empirical calibration determined
for THEMIS \cite{mcfadden08b}. The coefficients of this calibration,
however, vary from spacecraft to spacecraft and can slowly drift with
time. Unfortunately, the first epoch time for these coefficients was
in January 2008. Given the agreement in spacecraft potential observations
with radial distance for THC-THE (the only spacecraft for which EFI
was deployed shown in Figure~\ref{fig:profiles}a), we simply ensure
the inferred densities are consistent between spacecraft. The densities
for THD and THE agreed very well, however, THC exhibited some systematic
differences in density (Figure~\ref{fig:profiles}b). These differences
largely occurred at much smaller L-shells, nonetheless, we neglect
THC density observations for this reason.

To arrive at a radial density profile, we bin the spacecraft potential
inferred densities from THD and THE by radial distance using $0.1\,\mathrm{R_{E}}$
bins, taking the average. The results were subsequently median filtered
over $0.5\,\mathrm{R_{E}}$ and the profile was extended to the model
magnetopause \cite{shue98} using a constant extrapolation.

\section*{Data Availability}

THEMIS data and analysis software (SPEDAS) are available at \href{http://themis.ssl.berkeley.edu}{http://themis.ssl.berkeley.edu}.
The OMNI data was obtained from the NASA/GSFC OMNIWeb interface at
\href{http://omniweb.gsfc.nasa.gov}{http://omniweb.gsfc.nasa.gov}.
Wind data was obtained from the NASA/GSFC CDAweb interface \href{http://cdaweb.sci.gsfc.nasa.gov}{http://cdaweb.sci.gsfc.nasa.gov}.

\section*{Author Contributions}

M.O.A., H.H. and F.P. conceived of the study. M.O.A., H.H. and M.D.H.
performed analysis on the data. M.O.A. interpreted the results and
wrote the paper. V.A. gave technical support and conceptual advice.

\section*{Competing Interests}

The authors declare no competing interests.
\begin{acknowledgments}
We acknowledge valuable discussions within the International Space
Science Institute (ISSI), Bern, team 350 \textquotedblleft Jets downstream
of collisionless shocks\textquotedblright , led by F. Plaschke and
H. Hietala. We also thank D. Burgess for helpful discussions. H. Hietala
was supported by NASA NNX17AI45G and the Turku Collegium for Science
and Medicine. M.D. Hartinger was supported by NASA NNX17AD35G. We
acknowledge NASA contract NAS5-02099 for use of data from the THEMIS
Mission. Specifically K.~H. Glassmeier, U. Auster and W. Baumjohann
for the use of FGM data provided under the lead of the Technical University
of Braunschweig and with financial support through the German Ministry
for Economy and Technology and the German Center for Aviation and
Space (DLR) under contract 50~OC~0302; C.~W. Carlson and J.~P.
McFadden for use of ESA data; D. Larson and the late R.~P. Lin for
use of SST data; and J.~W. Bonnell and F.~S. Mozer for EFI data.
We acknowledge Wind plasma (courtesy of S. Bale and the late R.~P.
Lin) and magnetic field (courtesy of R. Lepping and A. Szabo) data.
We acknowledge Oleg Troshichev and the Department of Geophysics, Arctic
and Antarctic Research Institute for ground magnetometer data.
\end{acknowledgments}

\renewcommand{\bibsection}{\section{References}}
\bibliographystyle{unsrtnat}

\clearpage{}


\end{document}